\documentclass[twocolumn,showpacs,preprintnumbers,amsmath,amssymb,prl]{revtex4}

\usepackage{graphicx}
\usepackage{dcolumn}% Align table columns on decimal point

\usepackage[latin1]{inputenc}
\usepackage[T1]{fontenc}
\usepackage{ae,aecompl}

\begin{document}

\newcommand{\rar}{$\rightarrow$}
\newcommand{\lrar}{$\leftrightarrow$}

\newcommand{\beq}{\begin{equation}}
\newcommand{\eeq}{\end{equation}}
\newcommand{\bea}{\begin{eqnarray}}
\newcommand{\eea}{\end{eqnarray}}
\newcommand{\Req}[1]{Eq. (\ref{E#1})}
\newcommand{\req}[1]{(\ref{E#1})}
\newcommand{\degree}{$^{\rm\circ} $}
\newcommand{\pcite}{\protect\cite}
\newcommand{\pref}{\protect\ref}
\newcommand{\Rfg}[1]{Fig. \ref{F#1}}
\newcommand{\rfg}[1]{\ref{F#1}}
\newcommand{\Rtb}[1]{Table \ref{T#1}}
\newcommand{\rtb}[1]{\ref{T#1}}

\title{Comments on "Length scale dependence of DNA mechanical properties"}

\author{Alexey K. Mazur}
%\email{alexey@ibpc.fr}
\affiliation{UPR9080 CNRS, Université Paris Diderot, Sorbonne Paris Cité,\\
Institut de Biologie Physico-Chimique,\\
13, rue Pierre et Marie Curie, Paris,75005, France.}
%\\
%This line break forced with \textbackslash\textbackslash
 
%\date{\today}

%\baselineskip=0.8cm

%\begin{abstract}
%\end{abstract}%===================================================


\maketitle

Recent experimental data
\cite{Cloutier:04,Cloutier:05,Wiggins:06a,Mathew-Fenn:08b,Chen:10e,Vafabakhsh:12,Shi:13}
indicate that the elastic wormlike rod (WLR) model of DNA that works
well on long length scales may break down on shorter scales relevant
to biology. According to Noy and Golestanian (N\&G) \cite{Noy:12}
molecular dynamics (MD) simulations predict DNA rigidity close to
experimental data and confirm one scenario of such breakdown, namely,
that for lengths of a few helical turns, DNA dynamics exhibit
long-range bending and stretching correlations.  Earlier studies using
similar forcefields
\cite{Bruant:99,Lankas:00,Mzbj:06,Mzprl:10,Mzpre:11} concluded that
(i) MD systematically overestimate the DNA rigidity, and (ii) no
deviations from the WLR model are detectable \cite{Mzprl:07,Mzpre:09}.
Here it is argued that the data analysis in N\&G was incorrect and
that the earlier conclusions are valid.

Measuring DNA rigidity by MD requires rigorous analysis of statistical
errors. For high accuracy, trajectories must be several orders of
magnitude longer than the corresponding relaxation times, and optimal
conditions are met for dynamics of one helical turn
\cite{Mzjpc:08,Mzjpc:09,Mzprl:10,Mzpre:11}. Longer fragments are
difficult to study because the relaxation times for twisting and
bending grow with the DNA length as $L^2$ and $L^4$, respectively
\cite{Lankas:00}.

N\&G reported simulations of several DNA turns and implied that by
considering many internal stretches of long DNA one improves the
sampling. Unfortunately, this intuitive assertion is valid only for
stretching. For bending and twisting it fails and the sampling is even
reduced because internal fragments are not independent and the
spectrum of their relaxation times involves that of the whole DNA
\cite{Mzjpc:08,Mzjpc:09}. Besides, all relaxation times scale linearly
with the solvent viscosity \cite{Lankas:00} which is high for the
SPC/E water employed by N\&G \cite{Mark:01b,Mark:02}, therefore, the
overall accuracy of their quantitative estimates is undetermined.

Complex methods of analysis of all-atom DNA trajectories may hide
pitfalls. Before concluding that MD reveal a length-scale dependence
these methods should be validated on \underline{finite} WLR
trajectories. Notably, the only evidence of stretching correlations is
the convex plot of the variance of the end-to-end distance $V(L)$ when
the bending contribution is negligible \cite{Mathew-Fenn:08b}.  N\&G
estimated this contribution from angle fluctuations and subtracted it.
To validate this procedure one should check that it linearizes convex
$V(L)$ dependences for \underline{finite} WLR trajectories. This is
impossible, however, because, in \underline{finite} WLR ensembles,
distance and angle fluctuations may correspond to different apparent
persistence lengths.  Further, an isolated mode extracted by the
principal component analysis is correlated by construction. Such
correlations, therefore, represent a supporting evidence, only if a
similarly extracted WLR mode exhibits a different pattern, which was
not shown.

\begin{figure}[ht]
\centerline{\includegraphics[width=6cm]{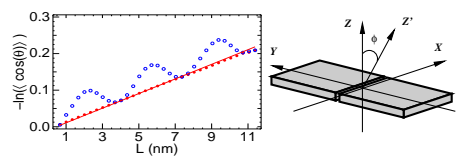}}
\caption{\label{Ffig1} Color online.
Length dependence of the bend angle $\theta$ for the WLR theory
(straight red line) and the Brownian dynamics of a discrete WLR model
\cite{Mzjpc:09}. A trajectory was analyzed with bend angle $\theta$
measured between either true Z-vectors (closed red circles) or
Z'-vectors biased by angle $\phi=10^\circ$ as shown on the right (open
blue circles). For integral numbers of helical turns the biasing
error is small.
}\end{figure}%=====================================================

Finally, oscillations with the helical period observed in some plots
are inevitably produced by the algorithm used for constructing
reference base-pair frames. This algorithm yields Cartesian frames
that fit the global helical axis in an ideal B-DNA. For MD structures
such frames are systematically biased. When this is taken into account
WLR dynamics produces periodical patterns similar to those shown in
N\&G (see \Rfg{fig1}). This is the main source of spurious
oscillations in MD data, and it also affects the analysis of
end-to-end distance fluctuations \cite{Mzbj:06}. However, it is an
artifact rather than an evidence of orientational memory, static
curvature, and so forth.

In summary, all-atom MD simulations with currently used forcefields
somewhat overestimate the rigidity of the double helix
\cite{Mzbj:06,Mzpre:11}, but agree with the WLR model \cite{Mzprl:07},
with no detectable correlations beyond a few base pair steps
\cite{Lankas:03,Mzpre:09}. It is possible that this agreement
disappears on somewhat longer time or length scales, or that new
effects will emerge with refined empirical forcefields. Additional
work is necessary to check this and also to clarify experimental
controversies \cite{Note1}.

\begin{acknowledgments}
\end{acknowledgments}%............................................

\bibliography{last}

\begin{thebibliography}{21}
\expandafter\ifx\csname natexlab\endcsname\relax\def\natexlab#1{#1}\fi
\expandafter\ifx\csname bibnamefont\endcsname\relax
  \def\bibnamefont#1{#1}\fi
\expandafter\ifx\csname bibfnamefont\endcsname\relax
  \def\bibfnamefont#1{#1}\fi
\expandafter\ifx\csname citenamefont\endcsname\relax
  \def\citenamefont#1{#1}\fi
\expandafter\ifx\csname url\endcsname\relax
  \def\url#1{\texttt{#1}}\fi
\expandafter\ifx\csname urlprefix\endcsname\relax\def\urlprefix{URL }\fi
\providecommand{\bibinfo}[2]{#2}
\providecommand{\eprint}[2][]{\url{#2}}

\bibitem[{\citenamefont{Cloutier and Widom}(2004)}]{Cloutier:04}
\bibinfo{author}{\bibfnamefont{T.~E.} \bibnamefont{Cloutier}} \bibnamefont{and}
  \bibinfo{author}{\bibfnamefont{J.}~\bibnamefont{Widom}},
  \bibinfo{journal}{Mol. Cell.} \textbf{\bibinfo{volume}{14}},
  \bibinfo{pages}{355} (\bibinfo{year}{2004}).

\bibitem[{\citenamefont{Cloutier and Widom}(2005)}]{Cloutier:05}
\bibinfo{author}{\bibfnamefont{T.~E.} \bibnamefont{Cloutier}} \bibnamefont{and}
  \bibinfo{author}{\bibfnamefont{J.}~\bibnamefont{Widom}},
  \bibinfo{journal}{Proc. Natl. Acad. Sci. U. S. A.}
  \textbf{\bibinfo{volume}{102}}, \bibinfo{pages}{3645} (\bibinfo{year}{2005}).

\bibitem[{\citenamefont{Wiggins et~al.}(2006)\citenamefont{Wiggins, Heijden,
  Moreno-Herrero, Spakowitz, Phillips, Widom, Dekker, and
  Nelson}}]{Wiggins:06a}
\bibinfo{author}{\bibfnamefont{P.~A.} \bibnamefont{Wiggins}},
  \bibinfo{author}{\bibfnamefont{T.~V.~D.} \bibnamefont{Heijden}},
  \bibinfo{author}{\bibfnamefont{F.}~\bibnamefont{Moreno-Herrero}},
  \bibinfo{author}{\bibfnamefont{A.}~\bibnamefont{Spakowitz}},
  \bibinfo{author}{\bibfnamefont{R.}~\bibnamefont{Phillips}},
  \bibinfo{author}{\bibfnamefont{J.}~\bibnamefont{Widom}},
  \bibinfo{author}{\bibfnamefont{C.}~\bibnamefont{Dekker}}, \bibnamefont{and}
  \bibinfo{author}{\bibfnamefont{P.~C.} \bibnamefont{Nelson}},
  \bibinfo{journal}{Nature nanotechnology} \textbf{\bibinfo{volume}{1}},
  \bibinfo{pages}{137} (\bibinfo{year}{2006}).

\bibitem[{\citenamefont{Mathew-Fenn et~al.}(2008)\citenamefont{Mathew-Fenn,
  Das, and Harbury}}]{Mathew-Fenn:08b}
\bibinfo{author}{\bibfnamefont{R.~S.} \bibnamefont{Mathew-Fenn}},
  \bibinfo{author}{\bibfnamefont{R.}~\bibnamefont{Das}}, \bibnamefont{and}
  \bibinfo{author}{\bibfnamefont{P.~A.~B.} \bibnamefont{Harbury}},
  \bibinfo{journal}{Science} \textbf{\bibinfo{volume}{322}},
  \bibinfo{pages}{446} (\bibinfo{year}{2008}).

\bibitem[{\citenamefont{Chen et~al.}(2010)\citenamefont{Chen, Fu, Zhou, and
  Yan}}]{Chen:10e}
\bibinfo{author}{\bibfnamefont{H.}~\bibnamefont{Chen}},
  \bibinfo{author}{\bibfnamefont{H.}~\bibnamefont{Fu}},
  \bibinfo{author}{\bibfnamefont{Z.}~\bibnamefont{Zhou}}, \bibnamefont{and}
  \bibinfo{author}{\bibfnamefont{J.}~\bibnamefont{Yan}}, \bibinfo{journal}{Int.
  J. Mod. Phys. B} \textbf{\bibinfo{volume}{24}}, \bibinfo{pages}{5475}
  (\bibinfo{year}{2010}).

\bibitem[{\citenamefont{Vafabakhsh and Ha}(2012)}]{Vafabakhsh:12}
\bibinfo{author}{\bibfnamefont{R.}~\bibnamefont{Vafabakhsh}} \bibnamefont{and}
  \bibinfo{author}{\bibfnamefont{T.}~\bibnamefont{Ha}},
  \bibinfo{journal}{Science} \textbf{\bibinfo{volume}{337}},
  \bibinfo{pages}{1097} (\bibinfo{year}{2012}).

\bibitem[{\citenamefont{Shi et~al.}(2013)\citenamefont{Shi, Herschlag, and
  Harbury}}]{Shi:13}
\bibinfo{author}{\bibfnamefont{X.}~\bibnamefont{Shi}},
  \bibinfo{author}{\bibfnamefont{D.}~\bibnamefont{Herschlag}},
  \bibnamefont{and} \bibinfo{author}{\bibfnamefont{P.~A.~B.}
  \bibnamefont{Harbury}}, \bibinfo{journal}{Proc. Natl. Acad. Sci. U. S. A.}
  \textbf{\bibinfo{volume}{110}}, \bibinfo{pages}{E1444}
  (\bibinfo{year}{2013}).

\bibitem[{\citenamefont{Noy and Golestanian}(2012)}]{Noy:12}
\bibinfo{author}{\bibfnamefont{A.}~\bibnamefont{Noy}} \bibnamefont{and}
  \bibinfo{author}{\bibfnamefont{R.}~\bibnamefont{Golestanian}},
  \bibinfo{journal}{Phys. Rev. Lett.} \textbf{\bibinfo{volume}{109}},
  \bibinfo{pages}{228101} (\bibinfo{year}{2012}).

\bibitem[{\citenamefont{Bruant et~al.}(1999)\citenamefont{Bruant, Flatters,
  Lavery, and Genest}}]{Bruant:99}
\bibinfo{author}{\bibfnamefont{N.}~\bibnamefont{Bruant}},
  \bibinfo{author}{\bibfnamefont{D.}~\bibnamefont{Flatters}},
  \bibinfo{author}{\bibfnamefont{R.}~\bibnamefont{Lavery}}, \bibnamefont{and}
  \bibinfo{author}{\bibfnamefont{D.}~\bibnamefont{Genest}},
  \bibinfo{journal}{Biophys. J.} \textbf{\bibinfo{volume}{77}},
  \bibinfo{pages}{2366} (\bibinfo{year}{1999}).

\bibitem[{\citenamefont{Lankas et~al.}(2000)\citenamefont{Lankas, Sponer,
  Hobza, and Langowski}}]{Lankas:00}
\bibinfo{author}{\bibfnamefont{F.}~\bibnamefont{Lankas}},
  \bibinfo{author}{\bibfnamefont{J.}~\bibnamefont{Sponer}},
  \bibinfo{author}{\bibfnamefont{P.}~\bibnamefont{Hobza}}, \bibnamefont{and}
  \bibinfo{author}{\bibfnamefont{J.}~\bibnamefont{Langowski}},
  \bibinfo{journal}{J. Mol. Biol.} \textbf{\bibinfo{volume}{299}},
  \bibinfo{pages}{695} (\bibinfo{year}{2000}).

\bibitem[{\citenamefont{Mazur}(2006)}]{Mzbj:06}
\bibinfo{author}{\bibfnamefont{A.~K.} \bibnamefont{Mazur}},
  \bibinfo{journal}{Biophys. J.} \textbf{\bibinfo{volume}{91}},
  \bibinfo{pages}{4507} (\bibinfo{year}{2006}).

\bibitem[{\citenamefont{Mazur}(2010)}]{Mzprl:10}
\bibinfo{author}{\bibfnamefont{A.~K.} \bibnamefont{Mazur}},
  \bibinfo{journal}{Phys. Rev. Lett.} \textbf{\bibinfo{volume}{105}},
  \bibinfo{pages}{018102} (\bibinfo{year}{2010}).

\bibitem[{\citenamefont{Mazur}(2011)}]{Mzpre:11}
\bibinfo{author}{\bibfnamefont{A.~K.} \bibnamefont{Mazur}},
  \bibinfo{journal}{Phys. Rev. E} \textbf{\bibinfo{volume}{84}},
  \bibinfo{pages}{021903} (\bibinfo{year}{2011}).

\bibitem[{\citenamefont{Mazur}(2007)}]{Mzprl:07}
\bibinfo{author}{\bibfnamefont{A.~K.} \bibnamefont{Mazur}},
  \bibinfo{journal}{Phys. Rev. Lett.} \textbf{\bibinfo{volume}{98}},
  \bibinfo{pages}{218102} (\bibinfo{year}{2007}).

\bibitem[{\citenamefont{Mazur}(2009{\natexlab{a}})}]{Mzpre:09}
\bibinfo{author}{\bibfnamefont{A.~K.} \bibnamefont{Mazur}},
  \bibinfo{journal}{Phys. Rev. E} \textbf{\bibinfo{volume}{80}},
  \bibinfo{pages}{010901} (\bibinfo{year}{2009}{\natexlab{a}}).

\bibitem[{\citenamefont{Mazur}(2008)}]{Mzjpc:08}
\bibinfo{author}{\bibfnamefont{A.~K.} \bibnamefont{Mazur}},
  \bibinfo{journal}{J. Phys. Chem. B} \textbf{\bibinfo{volume}{112}},
  \bibinfo{pages}{4975} (\bibinfo{year}{2008}).

\bibitem[{\citenamefont{Mazur}(2009{\natexlab{b}})}]{Mzjpc:09}
\bibinfo{author}{\bibfnamefont{A.~K.} \bibnamefont{Mazur}},
  \bibinfo{journal}{J. Phys. Chem. B} \textbf{\bibinfo{volume}{113}},
  \bibinfo{pages}{2077} (\bibinfo{year}{2009}{\natexlab{b}}).

\bibitem[{\citenamefont{Mark and Nilsson}(2001)}]{Mark:01b}
\bibinfo{author}{\bibfnamefont{P.}~\bibnamefont{Mark}} \bibnamefont{and}
  \bibinfo{author}{\bibfnamefont{L.}~\bibnamefont{Nilsson}},
  \bibinfo{journal}{J. Phys. Chem. B} \textbf{\bibinfo{volume}{105}},
  \bibinfo{pages}{9954} (\bibinfo{year}{2001}).

\bibitem[{\citenamefont{Mark and Nilsson}(2002)}]{Mark:02}
\bibinfo{author}{\bibfnamefont{P.}~\bibnamefont{Mark}} \bibnamefont{and}
  \bibinfo{author}{\bibfnamefont{L.}~\bibnamefont{Nilsson}},
  \bibinfo{journal}{J. Phys. Chem. B} \textbf{\bibinfo{volume}{106}},
  \bibinfo{pages}{9440} (\bibinfo{year}{2002}).

\bibitem[{\citenamefont{Lankas et~al.}(2003)\citenamefont{Lankas, Sponer,
  Langowski, and Cheatham}}]{Lankas:03}
\bibinfo{author}{\bibfnamefont{F.}~\bibnamefont{Lankas}},
  \bibinfo{author}{\bibfnamefont{J.}~\bibnamefont{Sponer}},
  \bibinfo{author}{\bibfnamefont{J.}~\bibnamefont{Langowski}},
  \bibnamefont{and} \bibinfo{author}{\bibfnamefont{T.~E.}
  \bibnamefont{Cheatham}, \bibfnamefont{III}}, \bibinfo{journal}{Biophys. J.}
  \textbf{\bibinfo{volume}{85}}, \bibinfo{pages}{2872} (\bibinfo{year}{2003}).

\bibitem[{Not()}]{Note1}
\bibinfo{note}{See discussion in Refs. 14-15, and also preprint
  arXiv:0904.2678.}

\end{thebibliography}

\end{document}